\documentclass[12pt,preprint]{aastex}
\usepackage{natbib}
\def\m{$^{\rm m}$}

\begin{document}

\title{UV dust attenuation in normal star forming galaxies:\\
 I. Estimating the $L_{TIR}/L_{FUV}$ ratio}
\author{ L. Cortese\altaffilmark{1,2}, A. Boselli\altaffilmark{1}, V. Buat\altaffilmark{1},
G. Gavazzi\altaffilmark{2}, S. Boissier\altaffilmark{3}, A. Gil de Paz\altaffilmark{3}, 
M. Seibert\altaffilmark{4}, B. F. Madore\altaffilmark{3,5}, D. C. Martin\altaffilmark{4}
%L. Cortese\altaffilmark{1}, A. Boselli\altaffilmark{1},
%G. Gavazzi\altaffilmark{2}, J. Iglesias-Paramo\altaffilmark{1},
%B. F. Madore\altaffilmark{3,4},
%T. Barlow\altaffilmark{5}, L. Bianchi\altaffilmark{6},
%Y.-I. Byun\altaffilmark{7}, J. Donas\altaffilmark{1},
%K. Forster\altaffilmark{5}, P. G. Friedman\altaffilmark{5},
%T. M. Heckman\altaffilmark{8}, P. Jelinsky\altaffilmark{9},
%Y.-W. Lee\altaffilmark{7},
%R. Malina\altaffilmark{1},
%D. C. Martin\altaffilmark{5}, B. Milliard\altaffilmark{1},
%P. Morrissey\altaffilmark{5}, S. Neff\altaffilmark{10},
%R. M. Rich\altaffilmark{11}, D. Schiminovich\altaffilmark{5},
%O. Siegmund\altaffilmark{9}, T. Small\altaffilmark{5},
%%K. Y. Sukyoung\altaffilmark{12},
%A. S. Szalay\altaffilmark{8},
%, M. A. Treyer\altaffilmark{5},
%B. Welsh\altaffilmark{9}, T. K. Wyder\altaffilmark{5}
}
\altaffiltext{1}{Laboratoire d'Astrophysique de Marseille, BP8, Traverse du Siphon, F-13376 Marseille, France}
\altaffiltext{2}{Universit\`a degli Studi di Milano - Bicocca, P.zza della Scienza 3,
20126 Milano, Italy}
\altaffiltext{3}{Observatories of the Carnegie Institution of Washington,
813 Santa Barbara St., Pasadena, CA 91101}
\altaffiltext{4}{California Institute of Technology, MC 405-47, 1200 East
California Boulevard, Pasadena, CA 91125}
\altaffiltext{5}{NASA/IPAC Extragalactic Database, California Institute
of Technology, Mail Code 100-22, 770 S. Wilson Ave., Pasadena, CA 91125}
%\altaffiltext{6}{Center for Astrophysical Sciences, The Johns Hopkins
%University, 3400 N. Charles St., Baltimore, MD 21218}
%\altaffiltext{7}{Center for Space Astrophysics, Yonsei University, Seoul
%120-749, Korea}
%\altaffiltext{8}{Department of Physics and Astronomy, The Johns Hopkins
%University, Homewood Campus, Baltimore, MD 21218}
%\altaffiltext{9}{Space Sciences Laboratory, University of California at
%Berkeley, 601 Campbell Hall, Berkeley, CA 94720}
%\altaffiltext{10}{Laboratory for Astronomy and Solar Physics, NASA Goddard
%Space Flight Center, Greenbelt, MD 20771}
%\altaffiltext{11}{Department of Physics and Astronomy, University of
%California, Los Angeles, CA 90095}
%\altaffiltext{12}{Oxford University, Astrophysics, Oxford OX1 3RH, United Kingdom}

\begin{abstract}
We analyze the dust attenuation properties of a volume-limited, optically-selected 
sample of normal star forming galaxies in nearby clusters as observed by GALEX.
The internal attenuation is estimated using three independent indicators, namely: the 
ratio of the total infrared to far-ultraviolet emission, the 
ultraviolet spectral slope $\beta$ and the Balmer decrement.
We confirm that normal galaxies follow a $L_{TIR}/L_{FUV}-\beta$ relation
offset from the one observed for starburst galaxies.
This offset is found to weakly correlate with the 
birthrate parameter, thus with the galaxy star formation history. 
We study the correlations of dust attenuation with other
global properties, such as the metallicity, dynamical mass, ionized gas attenuation, H$\alpha$ emission and 
mass surface density. 
Metal-rich, massive galaxies are, as expected, more heavily extinguished in
the UV than are small systems. 
For the same gas metallicity normal galaxies 
have lower $L_{TIR}/L_{FUV}$ ratio than starbursts, 
in agreement with the difference observed in the $L_{TIR}/L_{FUV}-\beta$ relation.
Unexpectedly we find however that normal star forming galaxies 
follow exactly the same relationship between metallicity and 
ultraviolet spectral slope $\beta$ determined for starbursts,
complicating our understanding of dust properties. 
This result might indicate a different dust geometry between 
normal galaxies and starbursts, but it could also be due 
to aperture effects eventually present in the IUE 
starbursts dataset.\\ 
The present multiwavelength study allows us to provide some empirical relations 
from which the total infrared to far ultraviolet ratio ($L_{TIR}/L_{FUV}$)
can be estimated when far infrared data are absent.
\end{abstract}

\section{Introduction}
\setcounter{footnote}{0}
The presence of dust in galaxies represents one
of the major obstacles complicating a direct quantification of the star formation activity
in local and high redshift galaxies.
Absorption by dust grains reddens the spectra at short wavelengths 
and modifies altogether the spectral energy distribution of galaxies.
Since the UV radiation is emitted by young stars 
($t<10^{8}$ yr) that are generally more affected by attenuation from surrounding dust clouds 
than older stellar 
populations, rest-frame UV observations can lead to incomplete and/or biased reconstructions 
of the star formation activity and star formation history of 
galaxies affected by dust absorption, unless proper corrections are applied.\\
In recent years our understanding of dust attenuation received a tremendous impulse from 
studies of local starburst galaxies 
(i.e.\citealp{calzetti94,heckman98,meurer99,calzetti01,charlot2000}), 
that were based on three indicators: 
the ratio of the total infrared to far-ultraviolet emission ($L_{TIR}/L_{FUV}$), the 
ultraviolet spectral slope $\beta$ (determined from a power-law fit of the form $f\sim\lambda^{\beta}$ 
to the UV continuum spectrum in the range 1300 and 2600 $\rm\AA$, \citealp{calzetti94}) and the Balmer decrement.
The total-IR (TIR) to UV luminosity ratio method (i.e. \citealp{buat92,xu95,meurer95,meurer99}) is 
based on the assumption that a fraction of photons emitted by stars and gas are 
absorbed by the dust. The dust heats up and subsequently re-emits the energy in the mid- and far-infrared.
The amount of UV attenuation can thus be quantified by means of an energy balance. 
This method is considered the most reliable estimator of the dust attenuation in star-forming galaxies
because it is almost completely independent of the assumed extinction mechanisms
(i.e. dust/star geometry, extinction law, see \citealp{buat96,meurer99,gordon00,witt00}).
When the spectrum is dominated by a young stellar population 
the ultraviolet spectral slope $\beta$, is found to have a weak dependence
on metallicity, IMF, and star formation history \citep{leiterer95}. Thus the difference between the observed  
$\beta$ and the one predicted by models can be entirely ascribed to dust attenuation \citep{meurer99}. 
However in systems with no or mild star formation activity the UV spectral slope can be 
strongly contaminated by the old stellar populations, whose contribution increases
$\beta$ (flattens the UV continuum, \citealp{samM83}). Thus the spectral slope of 
mildly star forming systems could be intrinsically 
different from the one of starburst galaxies, even in the absence of dust attenuation \citep{kong04}.\\
\cite{meurer99} have shown that in starburst galaxies the total far-infrared to ultraviolet luminosity ratio 
correlates with the ultraviolet spectral slope,
$\beta$ (commonly referred to as the IRX-UV relation).
They pointed out that this relation  
allows reliable estimates of the attenuation 
by dust at ultraviolet wavelengths based on $\beta$.\\
The Balmer decrement gives an estimate 
of the attenuation of ionized gas and not of the stellar continuum as in the previous two methods.
It is based on the comparison of 
the observed H$\alpha$/H$\beta$ ratio with its predicted value 
(2.86 for case B recombination, assuming an electronic density $n_{e}\leq10^{4}\rm ~cm^{-3}$ and temperature 
$\sim 10^{4}\rm ~K$; e.g., \citealp{osterb89}).
\cite{calzetti94} found a significant correlation 
between the ultraviolet spectral slope $\beta$ and the Balmer decrement H$\alpha$/H$\beta$.
Starting from this empirical relation they obtained an
attenuation law (known as the Calzetti attenuation law) often adopted 
to correct UV observations for dust attenuation in absence of both far-infrared observations and estimates 
of the ultraviolet spectral slope \citep{steidel99,glaze99}.\\
Unfortunately the above empirical relations have been established
only for starburst galaxies and they seem not to hold for normal star forming galaxies.
Recently, \cite{bell02} suggested that quiescent galaxies deviate
from the IRX-UV relation of starburst galaxies, because they tend to have
redder ultraviolet spectra at fixed total far-infrared to ultraviolet luminosity ratio. 
\cite{kong04} confirmed this result and interpreted the different behaviour of starbursts and normal galaxies 
as due to a difference in the star formation histories. 
They proposed that the offset from the starburst 
IRX-UV relation can be predicted using the birthrate parameter $b$ (e.g. the ratio of the
current to the mean past star formation activity).
However an independent observational confirmation of the  
correlation between the distance from the starburst IRX-UV relation and the 
birthrate parameter has not been obtained so far (\citealp{seibert05}).
Even the Calzetti law does not seem to be universal. 
\cite{buat02} showed that for normal star forming galaxies 
the attenuation derived from the Calzetti law is $\sim$0.6 mag
larger than the
one computed from the $F_{\rm FIR}/F_{\rm UV}$ ratio and their result 
has been recently confirmed by \cite{laird05}.\\
Why do normal star-forming galaxies behave differently from starbursts?
Do normal galaxies follow different empirical relations that can be exploited
to correct for dust attenuation in absence of far infrared observations?
If this is the case, is there a transition between starburst and normal galaxies?
Which physical parameters drive it? 
Answering these questions will be important for a better
understanding of the interaction of dust and radiation specifically in
nearby dusty star forming galaxies, but it also has direct
consequences for our understanding and interpretation of galaxy
evolution in a general context.
Firstly it seems mandatory to characterize the dust attenuation properties 
of normal galaxies, to compare them with the ones of starbursts and to derive new recipes for the
UV dust attenuation correction. 
This topic came once again to the fore with the launch of the \emph{Galaxy Evolution Explorer} (GALEX).
This satellite is delivering to the community an unprecedented amount of UV data 
on local and high redshift galaxies that require corrections for 
dust attenuation but currently lack far-infrared rest-frame data.
The time is ripe to explore new methods for correction of these data, 
that might provide new insights on galaxy evolution.
Whenever they can be combined with other data, GALEX observations provide the best available ultraviolet data
 for studying the dust attenuation properties of galaxies.
Multiwavelength photometric and spectroscopic 
observations are in fact mandatory in order to: determine metallicity, 
ionized gas attenuation ($A(H\alpha$)), luminosity and mass, 
test the validity of the relations followed by starbursts \citep{heckman98}, explore 
relations that might prove useful to correct ultraviolet magnitudes and to compare them 
with various models of dust attenuation.
Recent extensive spectroscopic and photometric surveys, like the Sloan Digital 
Sky Survey (SDSS, \citealp{sloan}) and the Two Degree Field Galaxy Redshift Survey (2dF, \citealp{2df})
have opened the path to studies of fundamental physical parameters based on
enormous datasets.
However, spectroscopic observations of nearby galaxies suffer from strong aperture effects, 
making these datasets not ideal for the purpose of the present investigation. 
In fact, \cite{jansen05} have recently shown that aperture effects produce both systematic and 
random errors on the estimate of star-formation, metallicity and attenuation. 
To reduce at least the systematic effects they suggest selecting only 
samples with fibres that capture $>$ 20\% of the light.
This requires $z>0.04$ and $z>0.06$ for SDSS and 2dF respectively: too distant to 
detect both giant and dwarf star forming systems with GALEX and IRAS.\\
Although significantly smaller than the SDSS, the dataset we have been building up
over the last 10 years
with data taken over a large stretch of the electromagnetic spectrum for 
few thousand galaxies in the local universe 
(worldwide available from the site GOLDMine; \citealp{goldmine}) 
turns out to be appropriate for the purposes of the present investigation.
It includes drift-scan mode integrated spectra, 
narrow band H$\alpha$ and broad band optical and near-infrared imaging for 
a volume limited sample of nearby galaxies in and outside rich clusters.
The combination of GALEX and IRAS observations with these ancillary data 
allows us to study the dust attenuation properties in a sizable sample of
normal star forming galaxies not suffering from the aperture bias
and to compare observations with model predictions.

In this first paper we investigate the relations between dust attenuation and global 
galaxy properties and compare them with the ones observed in starburst galaxies.
The aim of this work is to provide some empirical relations based  
on observable quantities (thus model independent) suitable for deriving
dust attenuation corrections when far infrared data are not available. 
For this reason all relations 
obtained throughout this paper will be given as a function of $L_{TIR}/L_{FUV}$, 
the observable that we consider the best dust attenuation indicator. 
We choose not to transform $L_{TIR}/L_{FUV}$ into a (model dependent) 
estimate of the far ultraviolet extinction $A(FUV)$, 
leaving the reader free to choose his/her preferred dust model  
(i.e. \citealp{meurer99,buat99,buat02,buat05,gordon00,panuzzo03,denis05a}, Inoue et al. in preparation).
Throughout the paper we assume that quantities are related linearly and 
residual plots are presented in order to test the validity of this hypothesis. 
Moreover, since we are looking for new recipies to estimate the $L_{TIR}/L_{FUV}$ ratio, 
this quantity has to be considered as the dependent variable, 
implying the use of an unweighted simple linear fit 
to estimate the best fitting parameters \citep{linear}.\\ 
A forthcoming paper will be focused on the comparison between models and observations 
hoping to gain a better understanding on the physics of dust attenuation and to study new recipes 
useful to convert $L_{TIR}/L_{FUV}$  into $A(FUV)$. 

\section{The Data}
\subsection{The optically-selected sample}
The analysis presented in this work is based on an optically selected sample of 
late-type galaxies (later than S0a) including 
giant and dwarf systems extracted from the Virgo Cluster Catalogue (VCC, \citealp{vcc}) 
and from the CGCG catalogue \citep{ZWHE61}.
The data include $\sim300$ square degrees covering
most of the Virgo, Abell1367 and Abell262 clusters,
the southwest part of the Coma cluster and part of the Coma-A1367 supercluster
(11h30m $<$ R.A. $<$ 13h30m; $\rm 18^{\circ} < decl.< 32^{\circ}$) observed in spring 2004 
as part of the All-sky Imaging Survey (AIS) and of the Nearby Galaxy Survey (NGS) carried out by GALEX 
in two UV bands: FUV ($\rm \lambda_{eff}=1530\AA, \Delta \lambda=400\AA$) 
and NUV ($\rm \lambda_{eff}=2310\AA, \Delta \lambda=1000\AA$).
Details of the GALEX instrument and characteristics can be found in 
\cite{martin05} and \cite{morrisey05}.
Our sample has the quality of being selected with the criterion of optical completeness.
All galaxies brighter than a threshold magnitude are selected in all areas.
In Coma-A1367 supercluster and A262 cluster all galaxies brighter than $m_{p}$=15.7 were selected
from the CGCG catalogue \citep{ZWHE61}. The Virgo region contains all galaxies brighter
than $m_{p}$=18 from the VCC catalogue \citep{vcc}.
We thus consider our sample an optically selected, volume limited sample.\\
We include in our analysis all late-type galaxies, 
detected in both NUV and FUV GALEX bands and in 
both 60 $\mu m$ and 100 $\mu m$ IRAS bands (157 objects).
Whenever available, we extracted UV fluxes from the deep NGS images, obtained with 
a mean integration time
of $\sim$ 1500 sec, complete to $m_{AB}$ $\sim$ 21.5 in the NUV and FUV. 
Elsewhere UV fluxes have been extracted from the shallower 
AIS images ($\sim$ 70 sq. degrees), obtained with a mean integration time of $\sim$ 100 sec, 
complete to $m_{AB}$ $\sim$ 20 in both the FUV and NUV bands. 
All UV images come from the Internal Data Release v1 (IR1.0). 
UV fluxes were obtained by integrating GALEX images
within elliptical annuli of increasing diameter up to the optical B band 25 mag 
arcsec$^{-2}$ isophotal radii, consistently
with the optical and near-IR images. Independent measurements of the same 
galaxies obtained in different 
exposures give consistent photometric results within 10 \% in the NUV and 15\% 
in the FUV in the AIS, and  a factor of $\sim$ two better for bright (NUV $\leq$16) galaxies. 
The uncertainty in the UV photometry 
is on average a factor of $\sim$ 2 better in the NGS than in the AIS, particularly
for faint objects. The typical uncertainty in the IRAS data is 15\% \citep{bosellised}.\\
UV and far-infrared data have been combined to multifrequency data.
These are optical and near-IR H imaging (mostly from \citealp{gav00,gav05,bosellised}), optical drift-scan 
spectra (\citealp{gavspectra}; Gavazzi et al. in prep.) and H$\alpha$ imaging 
(\citealp{bosgav02,bosha2,catinella,gavha,jorge}; Gavazzi et al. in prep.), 
great part of which are available from the 
GOLDMine galaxy database \citep{goldmine} (http:\slash \slash goldmine.mib.infn.it).
From the 157 galaxies selected we exclude Active Galactic Nuclei (AGN). AGNs have been selected using either 
the classification provided by NED, if available, or by inspection to the integrated spectra of 
\cite{gavspectra}: we exclude galaxies with 
$\log([OIII]/H\beta)>0.61/(\log([NII]/H\alpha) -0.05) +1.3$ \citep{agn}. 
This criterion reduces the sample to 128 galaxies, spanning a range of 
six magnitudes in B band (-22$<M_{B}<$-16) and of three orders of magnitude in mass\footnote{
Computed using the relation between $L_{H}$ and $M$ by \cite{phenomen}} ($9<M<12\rm ~M_{\odot}$).
Unfortunately ancillary data are not available for all galaxies observed by GALEX, we 
thus further divided the data in two subsamples.
Sixty six galaxies in the \emph{primary sample} have all the necessary complementary
data (e.g. H$\alpha$ photometry, H$\alpha$/H$\beta$ ratio, metallicity, H-band photometry; 
see \citealp{gav00,gav02,gavha,gavspectra} for the selection criteria adopted in each survey). 
The remaining 62 galaxies form the \emph{secondary sample}. 
%{\bf (See \citealp{gav00,gav02,gavha,gavspectra} for the selection criteria of each survey)., it 
%is difficult to determine if any systematic bias affects the two samples. 
We cannot exclude a possible 
contamination of AGN in the \emph{secondary sample}, since no spectra are available for these objects. 
In all figures objects belonging to the \emph{primary sample} will be indicated with filled circles 
while the \emph{secondary sample} as empty circles.
Since only galaxies belonging to the \emph{primary sample} are present 
in all the plots analyzed in the presented work, all correlations will be 
quantified using only the \emph{primary sample}. 
Data from UV to near-IR have been corrected for Galactic extinction
according to \cite{burnstein82}.\\
We assume a distance of 17 Mpc for the members
of Virgo Cluster A, 22 Mpc for Virgo Cluster B, and 32 Mpc for objects
in the M and W clouds \citep{gav99}.
Members of the Cancer, A1367, and Coma clusters are assumed to
lie at distances of 65.2, 91.3, and 96 Mpc, respectively.
Isolated galaxies in the Coma supercluster are assumed 
at their redshift distance, adopting $H_{0}$ = 75 $\rm km~s^{-1}~Mpc^{-1}$.

\subsection{The starburst sample}
In order to compare the properties of 
our sample with starbursts, we compile a dataset of starburst galaxies observed 
by IUE from 
the sample of \cite{calzetti94}.
We consider 29 galaxies, excluding AGNs and galaxies that have not been observed by IRAS at 60 or 100 $\mu $m. 
Complementary data such as FIR, H$\alpha$ fluxes and Balmer decrements are taken from \cite{calzetti95}, 
metallicities come from \cite{heckman98} and 
H-band photometry (available only for 18 galaxies) from \citep{calzH}.
Excluding the far infrared fluxes, all these quantities 
are obtained within an apertures of $\sim20\times10 \rm arcsec^{2}$, consistent with 
IUE observations \cite{calzetti94}.
Thus we stress that aperture effects could strongly affect any comparison with normal galaxies 
for which all data are homogeneously integrated values. 
First of all, if the UV emission is more extended than IUE field of view 
the $L_{TIR}/L_{FUV}$ ratio is overestimated\footnote{However \cite{meurer99} 
argued that the majority of UV flux for their starburst sample lies within the 
IUE aperture}. In addition, even when physical quantities are obtained in the same 
IUE apertures, the presence of age and metallicity gradients in galaxies makes 
not trivial any comparison with the integrated values obtained for normal star forming 
galaxies \citep{jansen05}.
All the observables, but the ultraviolet spectra slope $\beta$, are calibrated in a 
consistent way with our sample of normal galaxy. 
The ultraviolet spectral slope of starbursts is obtained by fitting IUE spectra \citep{calzetti94}, 
while for GALEX observations 
it comes from the FUV-NUV color index (see next Section). However, as shown by \cite{kong04}, 
these two calibrations are consistent each other and do not introduce any systematic difference between 
the two samples.

\section{The $L_{TIR}/L_{FUV}-\beta$ relation for normal star-forming galaxies}
\label{parTIRFUV}
\begin{figure}
\epsscale{0.8}
\plotone{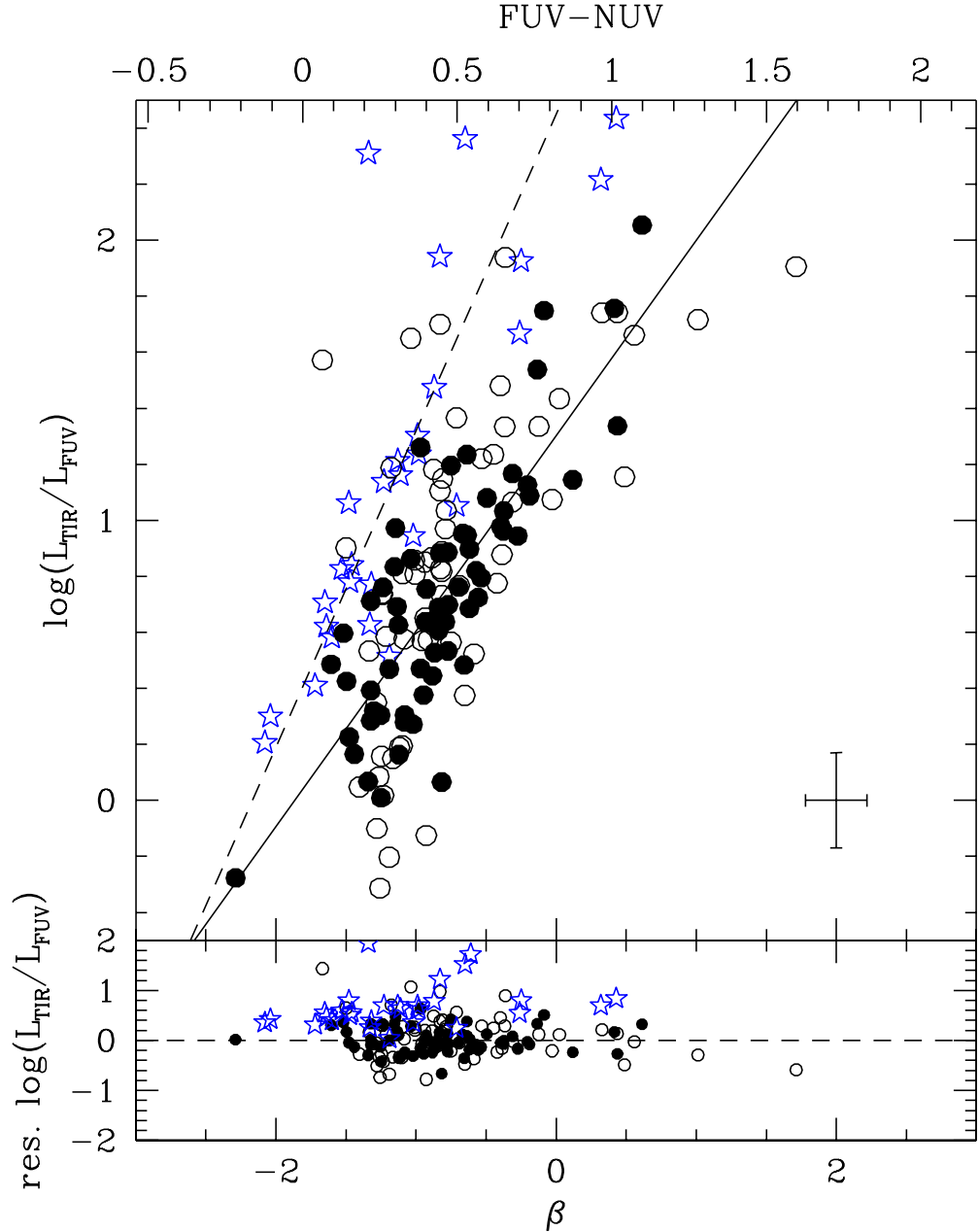}
\small{\caption{Ratio of the total infrared to far ultraviolet luminosity as a function of the ultraviolet spectral
slope (lower x-axis) and the FUV-NUV color (upper x-axis). Open circles indicates our \emph{secondary sample} while
filled circles represent the \emph{primary sample}. The dashed line represents the best linear fit to starburst IRX-UV relation. The
solid line indicates the best bisector linear fit for our \emph{primary sample}.
The stars indicate the sample of IUE starbursts.
Mean error bars for the plotted data are shown in the lower right corner,
in this and subsequent figures. The residuals from the best linear 
fit for normal galaxies are shown in the bottom panel.}
\label{IRX}}
\end{figure}

\cite{meurer99} have shown that the ratio of far infrared to far ultraviolet luminosity 
tightly correlates with the UV colors
of starburst galaxies. This relation, known as the infrared excess-ultraviolet (IRX-UV) relation,
is often presented as $\beta$ vs. $L_{TIR}/L_{FUV}$ relation.
As discussed in the introduction, we will 
refer throughout this paper to the $L_{TIR}/L_{FUV}$ ratio as the best indicator of UV dust attenuation and we will 
calibrate on it all the following relations. 
In order to determine the dust emission, we compute the total infrared flux emitted in the
range 1-1000 $\mu m$, following \cite{dale01}:
\begin{eqnarray}
 \label{tir}
 \log(f_{TIR})  =  \log(f_{FIR}) + 0.2738 - 0.0282\times\log(\frac{f_{60}}{f_{100}})+
 \nonumber\\
 +0.7281\times\log(\frac{f_{60}}{f_{100}})^{2}+0.6208\times\log(\frac{f_{60}}{f_{100}})^{3} +
 \nonumber\\
 + 0.9118\times\log(\frac{f_{60}}{f_{100}})^{4}
 \end{eqnarray}
where $f_{FIR}$ is the far-infrared flux, defined as the flux between 42 and 122 $\mu m$ \citep{helou88}:
\begin{equation}
 f_{FIR} = 1.26 \times (2.58 \times f_{60} + f_{100}) \times 10^{-14} ~~[\rm Wm^{-2}] 
 \end{equation}
and $f_{60}$ and $f_{100}$ are the IRAS fluxes measured at 60 and 100 $\mu m$ (in Jansky).
The total infrared luminosity is thus:
\begin{equation}
L_{TIR} = 4\pi D^{2} f_{TIR}
\end{equation}
The $\beta$ parameter as determined from GALEX colors
is very sensitive to the galaxy star formation history (see for example \citealp{calzetti05}).
For this reason  we assume throughout this paper $\beta$ as defined by \cite{kong04}:
\begin{eqnarray}
\beta = \frac{log(f_{FUV}) - log(f_{NUV})}{-0.182}=
\nonumber\\
 =2.201\times(FUV - NUV) -1.804
\end{eqnarray}
where $f_{FUV}$ and $f_{NUV}$ are the near and far ultraviolet observed fluxes respectively (in 
$\rm erg~ cm^{2} ~s^{-1}~ \AA^{-1}$), and 
FUV and NUV are the observed magnitudes.\\ 
The relationship between the ratio of total infrared luminosity ($L_{TIR}$) obtained from (\ref{tir})
to the far-ultraviolet fluxes and the UV spectral slope $\beta$ (or the FUV-NUV color) for our sample of
nearby star forming galaxies is given in Fig.\ref{IRX}. 
Several functional forms of the $L_{TIR}/L_{FUV}-\beta$ relation can be found in the literature
(i.e. \citealp{meurer99,kong04}); we simply adopt a linear fit:   
$\log(L_{TIR}/L_{FUV})=a\times\beta + b$. This functional form is 
consistent with other previously proposed for $\beta > -2$, while it diverges for $\beta<-2$. 
Since the majority of normal and starbursts galaxies have $\beta > -2$
our choice is justified. This represents the simplest and less parameter dependent way 
to study the relation between two quantities.\footnote{We tested this hypothesis fitting our data with functional 
forms similar to the ones proposed by \cite{meurer99} and \cite{kong04}: no significative improvement in the 
scatter of this relation is obtained.}
We find a strong correlation (Spearman correlation coefficient $r_{s}\sim$0.76 for the \emph{primary sample} and $r_{s}\sim$0.65 
for the \emph{secondary sample}, both corresponding to a probability $P(r_s)>$99.9\% that the two variables are correlated) 
between the total infrared to far ultraviolet ratio and the spectral slope, but significantly different from 
the one observed for starburst galaxies 
(dashed line in Fig.\ref{IRX}; \citealp{meurer99}). A $\chi^2$ test rejects at a confidence level higher 
than 99.9\%, that the two samples follow the same relation.
The best linear fit for our \emph{primary sample} (solid line in Fig.\ref{IRX}) is:
\begin{equation}
\label{irxsample}
\log(\frac{L_{TIR}}{L_{FUV}})= (0.70\pm0.06)\times\beta + (1.30\pm0.06)
\end{equation}
The uncertainty in the estimate of the $L_{TIR}/L_{FUV}$ using equation (\ref{irxsample})
is $\sim0.26\pm0.02$ dex for the \emph{primary sample} but it increases 
to $\sim0.35\pm0.03$ dex, if we consider the whole sample
(e.g.  \emph{primary} and \emph{secondary} samples), 
consistent with the mean uncertainty observed for starburst galaxies \citep{meurer99}.
A large contribution ($\sim0.21\pm0.02$ dex) to the observed scatter in Eq.(\ref{irxsample}) is 
due to the uncertainty on the estimate of $L_{TIR}/L_{FUV}$ and $\beta$.
This result confirms once more that the $L_{TIR}/L_{FUV}-\beta$ relation for normal galaxies deviates from the one
observed for starbursts, as pointed out
by previous studies of nearby galaxies 
(i.e. \citealp{bell02,kong04,samM83,buat05,seibert05,denis05a}, Boissier et al. in prep.) and individual
HII regions in nearby galaxies \citep{calzetti05}.

\subsection{The dependence on the birthrate parameter}
\begin{figure}
\epsscale{0.6}
\plotone{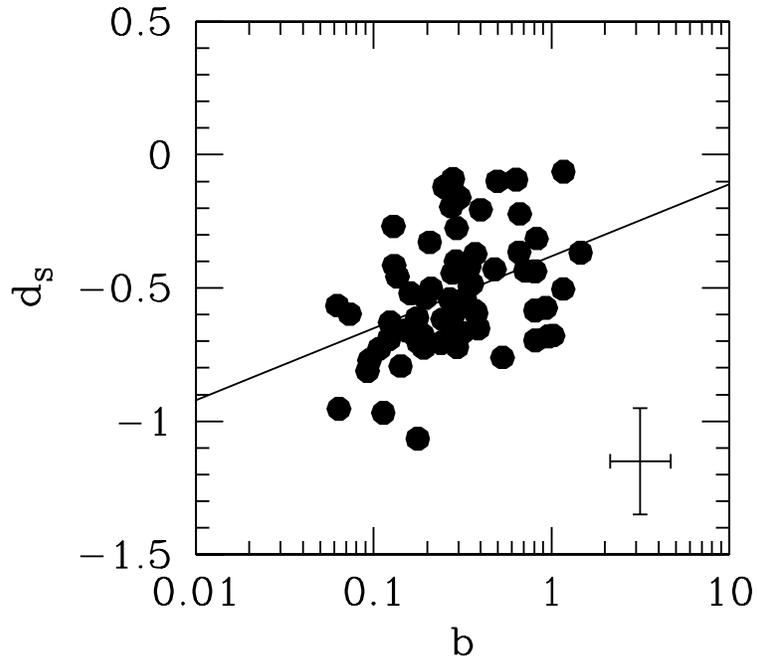}
\small{\caption{Relation between the birthrate parameter computed from the H$\alpha$ emission, and the distance 
from the $L_{TIR}/L_{FUV}-\beta$ relation for starbursts. The solid line represents the best linear fit.}
\label{resbHa}}
\end{figure}

What physical mechanisms drive the difference observed in the $L_{TIR}/L_{FUV}-\beta$ between normal star forming galaxies 
and starbursts? 
Recently \cite{kong04} interpreted the offset as an effect of the different star formation history experienced by
galaxies and proposed that the distance from the starburst IRX-UV can be predicted using the birthrate parameter $b$ 
(e.g. the ratio of the current to the mean past star formation activity, \citealp{kennbirth}).
In order to test if the perpendicular distance $d_{S}$ from the $L_{TIR}/L_{FUV}-\beta$ relation for 
starbursts correlates with the star formation history of normal galaxies, 
we compute the birthrate parameter following \cite{boselli}:
\begin{equation}
\label{b}
b = \frac{SFRt_{0}(1-R)}{L_{H}(M_{tot}/L_{H})(1-DM_{cont})}
\end{equation}
where $R$ is the fraction of gas that stellar winds re-injected into
the interstellar medium during their lifetime ($\sim$ 0.3, \citealp{kennbirth}), 
$t_{0}$ is the age of the galaxy (that we assume $\sim$12 Gyr),
$DM_{cont}$ is the dark matter contribution to the $M_{tot}/L_{H}$
ratio at the optical radius (assumed to be 0.5; \citealp{boselli}).
We compute the H-band luminosity following \cite{gav02}:
\begin{displaymath} 
\log L_{H} = 11.36 - 0.4\times H + 2\times\log(D) ~~~\rm [L_{\odot}]
\end{displaymath}
where $D$ is the distance to the source (in Mpc), and  
the SFR from the H$\alpha$ luminosity (corrected for $\rm[NII]$ contamination 
and for dust extinction 
using the Balmer decrement, see Appendix A) following \cite{boselli}:
\begin{equation}
\label{SFR}
SFR = \frac{L_{H\alpha}}{1.6 \times 10^{41}} ~~~\rm [M_{\odot}/yr]
\end{equation}
Fig.\ref{resbHa} shows the relation between the birthrate parameter (eq.\ref{b}) and the distance from the
$L_{TIR}/L_{FUV}-\beta$ relation for starburst galaxies.
The two quantities are correlated ($r_{s}\sim$0.40, corresponding to a correlation probability $P(r_s)\sim$99.8\%) but with a 
large scatter. Given the value of observational uncertainties, it is not worth trying to 
use the observed trend to reduce the dispersion in the $L_{TIR}/L_{FUV}-\beta$ relation 
for normal galaxies.
This result confirms that 
part of the dispersion in the $L_{TIR}/L_{FUV}-\beta$ relation
for normal star forming galaxies appears an effect of the different star formation history 
experienced by galaxies, as proposed by \cite{kong04}. 
%However, {\bf given the intrinsic uncertainties to the $\log(L_{TIR}/L_{FUV})$ and $\beta$ measurements, no significative} 
%improvement in the scatter of the $L_{TIR}/L_{FUV}-\beta$ relation
%is obtained when the residual dependences from 
%the star formation history are accounted for. 
%Given the uncertainty of $\sim$0.2 and $\sim$0.22 intrinsic to the $\log(L_{TIR}/L_{FUV})$ and $\beta$ measurement
%respectively, it is not worth trying other corrections.

\section{The $\beta$-$A(H\alpha$) relation}
\begin{figure}
\epsscale{0.8}
\plotone{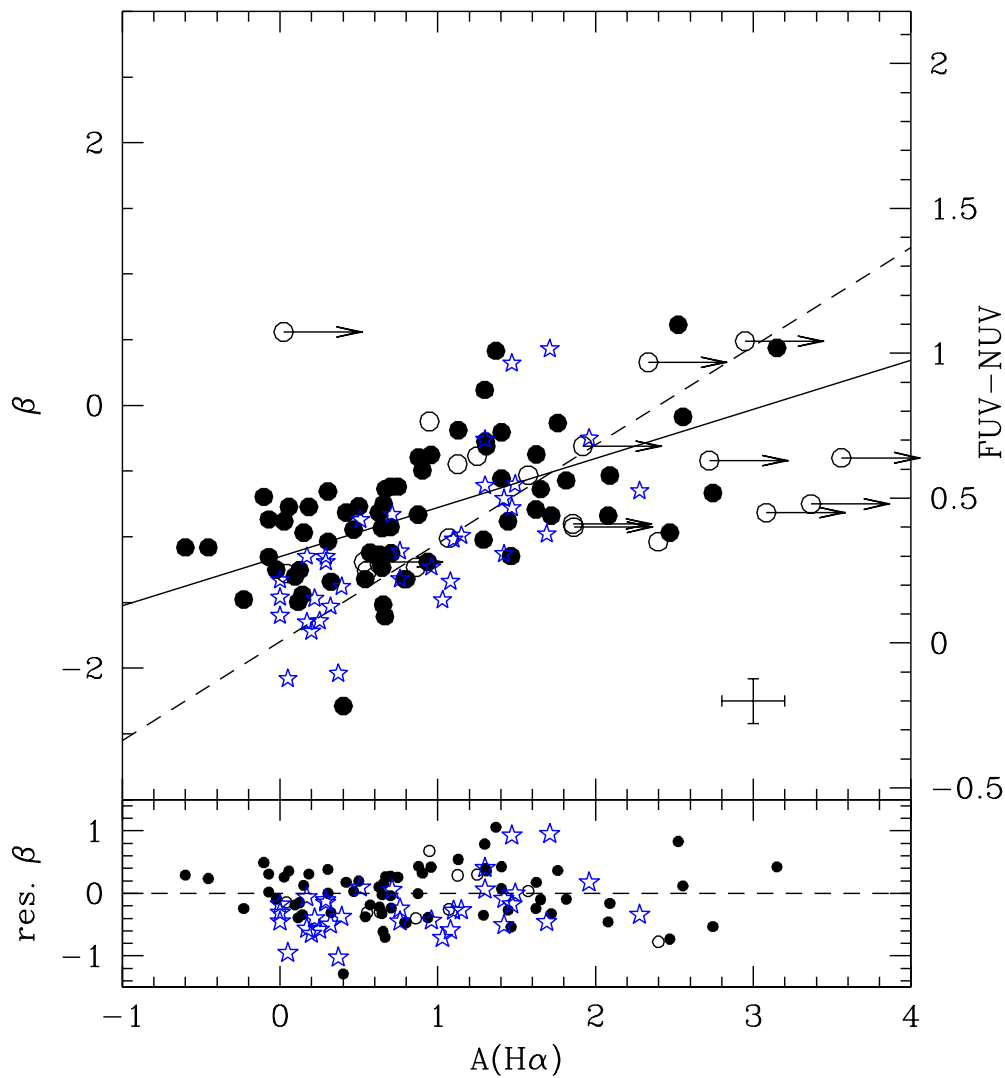}
\small{\caption{The relation between the ultraviolet spectral slope $\beta$ and the H$\alpha$ attenuation obtained
 from the Balmer decrement.
Symbols are as in Fig.\ref{IRX}. Solid line represents the best linear fit 
to our \emph{primary} sample (equation \ref{calznew}) while the dashed line indicate the
best-fit for starburst galaxies obtained by \cite{calzetti94} (equation \ref{calzlaw}). Arrows 
indicate galaxies for which the value of A(H$\alpha$) is a lower limit of the real value 
(i.e. H$\beta$ undetected). The residuals from the best linear 
fit for normal galaxies are shown in the bottom panel.}
\label{calzetti}}
\end{figure}

\cite{calzetti94} found a strong relationship between the ultraviolet spectral slope $\beta$ and
the Balmer decrement H$\alpha$/H$\beta$. For our starburst sample these two quantities 
are correlated ($r_{s}\sim$0.81) as follows (see also blue stars in Fig.\ref{calzetti}):
\begin{equation}
\label{calzlaw}
\beta = (0.75\pm0.10) \times  A(H\alpha) - (1.80\pm0.13)
\end{equation}
This empirical relation was used by \cite{calzetti94} 
to deduce an attenuation law (the Calzetti law), often applied to high redshift galaxies (i.e.
\citealp{steidel99,glaze99}).
Contrary to the $L_{TIR}/L_{FUV}-\beta$ relation the Calzetti law has not yet been tested 
for a sample of normal star forming galaxies.
\cite{buat02} showed that for normal star forming galaxies the attenuation derived 
from the Calzetti law is $\sim$0.6 larger than the one computed from $FIR/UV$ ratio.
This result has been recently confirmed by \cite{laird05} on
star forming galaxies at $z\sim$1.
In order to check the Calzetti law on our sample we use the measure of the 
H$\alpha$/H$\beta$ described in Appendix A.
Fig. \ref{calzetti} shows the relation between $\beta$ and $A(H\alpha$) for our sample (empty and 
filled circles). For the \emph{primary sample} we obtain $r_{s}\sim$0.58 ($P(r_s)>$99.9\%) and:
\begin{equation}
\label{calznew}
\beta = (0.37\pm0.07) \times  A(H\alpha) - (1.15\pm0.08)
\end{equation}
flatter than for starburst galaxies (see Fig.\ref{calzetti}).
At low $A(H\alpha)$ normal galaxies show on average a less steep 
ultraviolet spectral slope than starbursts. 
In addition normal galaxies with the same value of $\beta$ span a range of $\sim$ 1 mag in $A(H\alpha$).
At higher attenuation the two samples appear consistent.
Our result suggest that the Calzetti law cannot be applied to normal galaxies. 
On the contrary, the relation between $\beta$ and $A(H\alpha$) for normal galaxies, 
could be used to obtain a new attenuation law.

\section{Relations between dust attenuation indicators and global properties.}

\subsection{Metallicity}
\begin{figure}
\epsscale{0.8}
\plotone{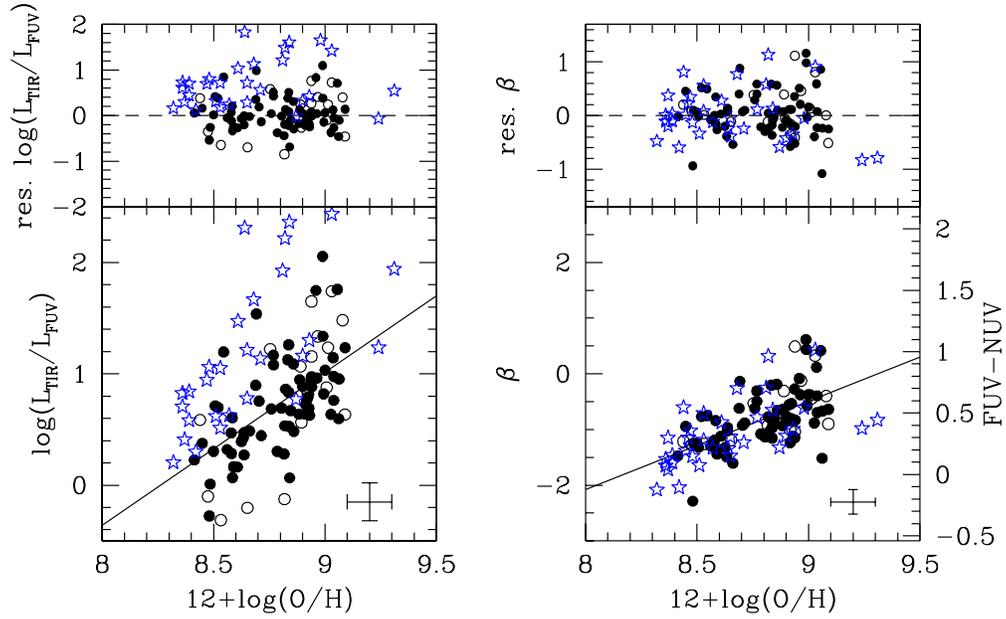}
\small{\caption{Relation between gas metallicity and the $L_{TIR}/L_{FUV}$ ratio (left) or $\beta$ (right). Symbols are as in Fig.\ref{IRX}.
The solid lines show the best linear fit for our \emph{primary} sample. The residuals from the best linear 
fits for normal galaxies are shown in the upper panels.}
\label{metal}}
\end{figure}

\begin{figure}
\epsscale{0.8}
\plotone{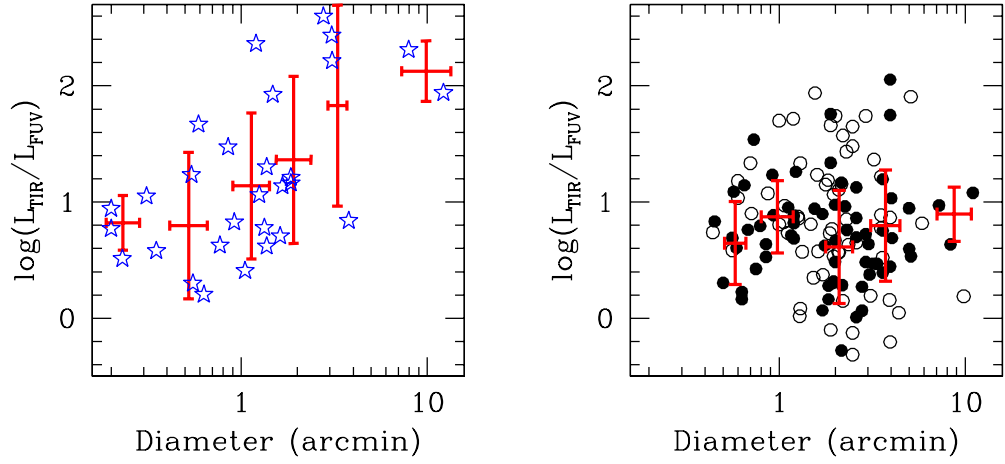}
\small{\caption{Relation between the galaxy size and the $L_{TIR}/L_{FUV}$ ratio for starburst (left panel) and 
normal galaxies (right panel). Symbols are as in Fig. \ref{IRX}. Mean values and uncertainties in bins of 0.30
$\log(Diameter)$ are given.}
\label{aperture}}
\end{figure}

\cite{heckman98} have shown that the ultraviolet spectral slope and
metallicity of starbursts are well correlated.
To determine the metal content of our galaxies
we average five different empirical determinations based on the following line ratios:
$\rm R_{23} \equiv ([OII]\lambda3727+[OIII]\lambda4959,5007)/H\beta$
\citep{zaritsky94,mcg91}, $\rm [NII]\lambda6583/[OII]\lambda3727$
\citep{kewley02}, $\rm [NII]\lambda6583/H\alpha$ \citep{vanzee98} and
$\rm [OIII]\lambda5007/ [NII]\lambda6583$ \citep{dutil99}.
The mean uncertainty in the abundances is $\rm 0.10 dex$.
In Fig. \ref{metal} we study the relationship between the gas metallicities and 
the $L_{TIR}/L_{FUV}$ ratio (left-panel) and $\beta$ (right-panel) for normal star forming 
and starburst galaxies.
For normal galaxies the $L_{TIR}/L_{FUV}$ ratio correlates ($r_{s}\sim$0.59, $P(r_s)>$99.9\%) with the gas abundance:
\begin{equation}
\label{relmet}
\log(\frac{L_{TIR}}{L_{FUV}}) = (1.37\pm0.24)\times 12+\log(O/H) - (11.36\pm2.11)
\end{equation}
with a dispersion of $\sim0.35\pm0.03$ in $\log(L_{TIR}/L_{FUV})$.
As for the $L_{TIR}/L_{FUV}-\beta$ relation normal galaxies differ from starbursts.
At comparable metallicity normal galaxies show a lower $L_{TIR}/L_{FUV}$ (lower attenuation) than
starbursts, in agreement with the recent result by \cite{samprof04} 
who studied radial extinction profiles of nearby late-type galaxies
using FOCA and IRAS observations.
Unexpectedly we find however that normal star forming galaxies 
follow exactly the same (significant, $r_{s}\sim0.58$,  $P(r_s)>$99.9\%) relationship between metallicity and 
ultraviolet spectral slope $\beta$ determined for starbursts by \cite{heckman98}
(see right panel of Fig.\ref{metal}).
This might indicate that even though a normal and a starburst galaxy 
with similar gas metallicity have similar 
UV spectral slopes, they suffer from a significantly different dust attenuation, 
perhaps suggesting a different dust geometry \citep{witt00}. 
However we stress that this effect might occur due 
to aperture effects in the IUE data: while $\beta$ is not significantly 
contaminated by aperture effects, the $L_{TIR}/L_{FUV}$ ratio could be overestimated 
producing the observed trend (the total infrared luminosity is obtained by 
integrating the IRAS counts over the full galaxy extension, while the ultraviolet one 
is taken from IUE's significantly smaller aperture $20\times10~\rm arcsec^{2}$).
%We do not find however a systematic trend between the offset of starburst from normal 
%galaxies (equation \ref{relmet}, Fig.\ref{metal} left panel) and the starburst diameter. 
This idea could be supported by the correlation ($r_{s}\sim0.49$, $P(r_s)>$99.9\% see Fig.\ref{aperture}) observed 
between the starbursts'  optical
diameters and the $L_{TIR}/L_{FUV}$ ratio, completely absent in our sample of normal galaxies 
($r_{s}\sim0.006$, $P(r_s)\sim$25\%). 
GALEX observations of starburst galaxies will rapidly solve this riddle.\\
We also checked the dependence of the dust attenuation on gas to dust ratio. 
Unfortunately, given the large errors on the estimate of the dust mass, 
we only obtain a weak relationship between the two quantities (see Appendix B).

\subsection{Luminosity}
\begin{figure}
\epsscale{0.8}
\plotone{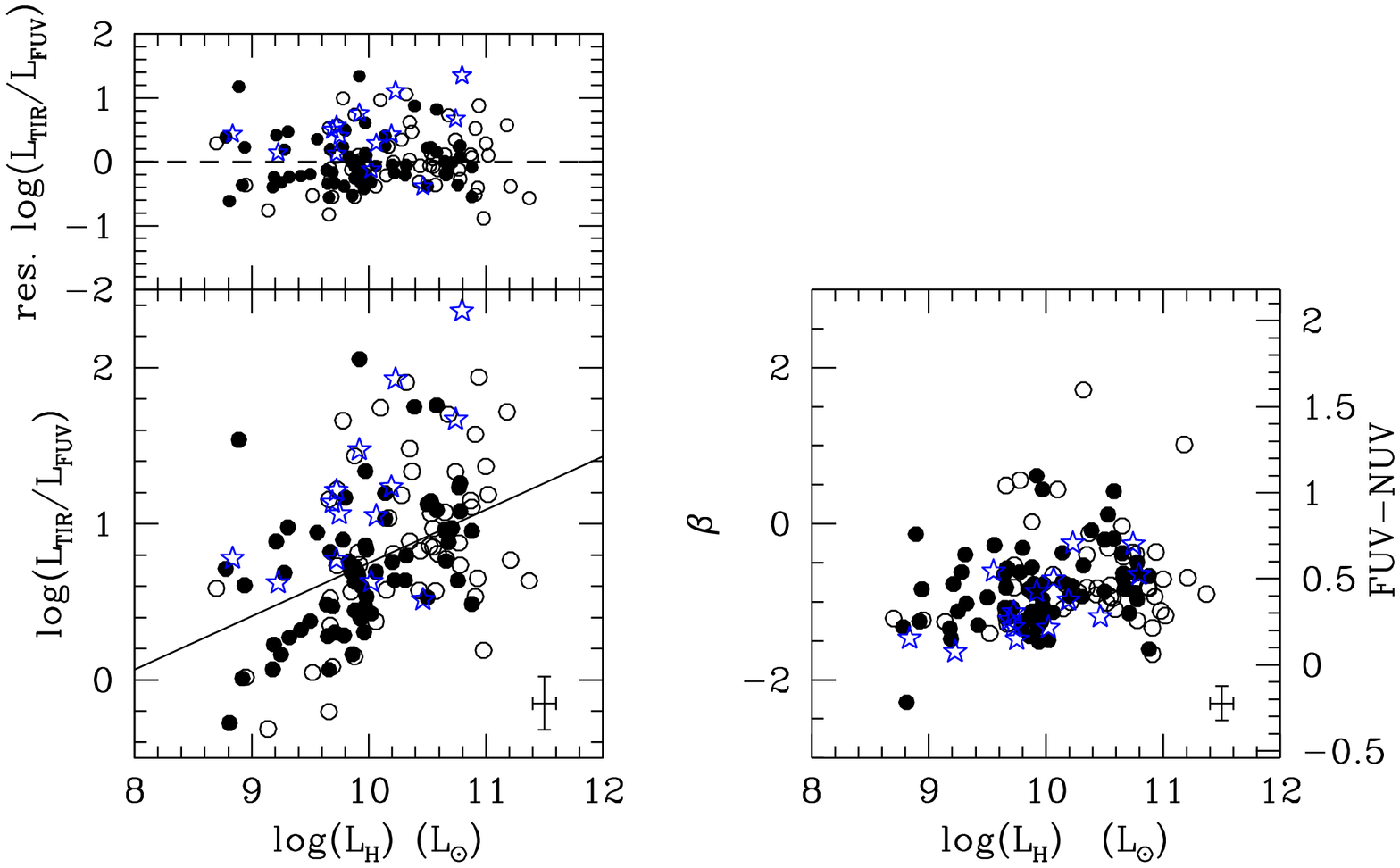}
\small{\caption{Relation between the H-band luminosity and the $L_{TIR}/L_{FUV}$ ratio (left) or $\beta$ (right). Symbols are as in Fig.
\ref{IRX}.
The solid line shows the best linear fit for our \emph{primary} sample. The residuals from the best linear 
fit for normal galaxies are shown in the upper panel.}
\label{hlum}}
\end{figure}

\begin{figure}
\epsscale{0.8}
\plotone{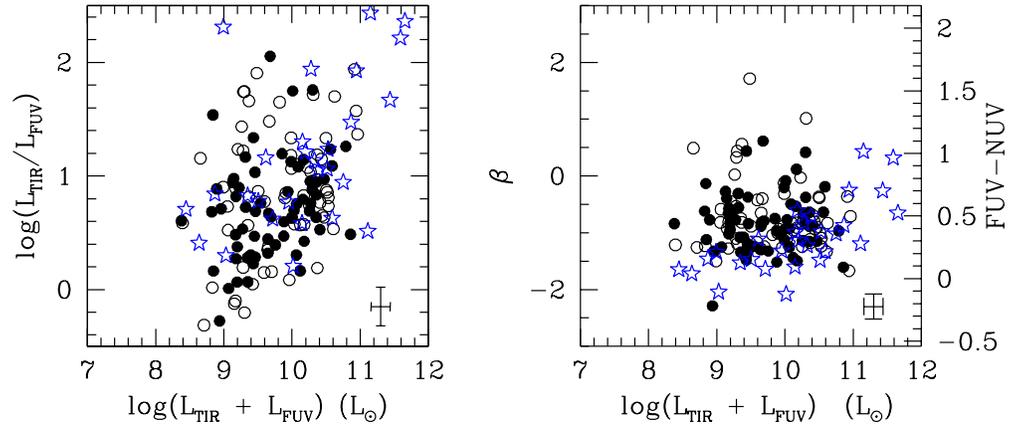}
\small{\caption{Relation between the TIR+FUV luminosity and the $L_{TIR}/L_{FUV}$ ratio (left) or $\beta$ (right). Symbols are as in Fig.
\ref{IRX}.
%The {\bf solid line} shows the best linear fit for our \emph{primary} sample.
}
\label{totlum}}
\end{figure}

Since it is well known that the metallicity of normal galaxies 
strongly correlates with galaxy luminosity 
(e.g. \citealp{skillman89,zaritsky94})
and mass (e.g. \citealp{tremonti04}), it is worth considering the correlation between attenuation 
and galaxy luminosity. Fig.\ref{hlum} shows 
the relationships between the dust attenuation indicators $L_{TIR}/L_{FUV}$ and $\beta$ and the 
H-band luminosity.
The infrared to far ultraviolet ratio correlates ($r_{s}\sim$0.49, $P(r_s)>$99.9\%) with the total H-band luminosity: 
\begin{equation}
\log(\frac{L_{TIR}}{L_{FUV}}) =  (0.34\pm0.10)\times \log(\frac{L_{H}}{L_{\odot}}) - (2.66\pm0.88)
\end{equation}
The dispersion of this relation is $\sim0.39\pm0.03$ in $\log(L_{TIR}/L_{FUV})$.
Since the H-band luminosity is proportional to the dynamical mass \citep{phenomen}, 
this implies a relationship between dust attenuation and dynamical mass. 
Also in starbursts the total H-band luminosity is correlated ($r_{s}\sim$0.37, $P(r_s)\sim$99.5\%) with the $L_{TIR}/L_{FUV}$ ratio 
and the great part of starbursts appear offset (to 99\% confidence level) from the relation of normal galaxies.
On the contrary, no difference is observed between the two samples in the $\beta$-$L_{H}$ plot, 
in agreement with what observed for metallicity.
Finally Fig.\ref{totlum} shows the relation between the bolometric luminosity ($\rm L_{TIR}+L_{FUV}$) and the dust attenuation, computed 
assuming that the UV emission is absorbed by dust and emitted in the far infrared.
The correlation coefficient  
($r_{s}\sim$0.31, $P(r_s)\sim$98\%) indicates that the two quantities correlate, 
as for starburst galaxies \citep{heckman98}. 
This is not the case if we examine the relation between the ultraviolet spectral slope $\beta$ and 
the bolometric luminosity (Fig.\ref{totlum} right panel): while there is no correlation 
($r_{s}\sim0.002$, $P(r_s)\sim$20\%) for our sample of normal 
galaxies, a clear relation ($r_{s}\sim0.68$, $P(r_s)>$99.9\%) holds for starbursts.
Starbursts with higher bolometric luminosity (high TIR emission) 
show lower ultraviolet slope, consistent with the idea that high TIR emission corresponds 
to high attenuation (low $\beta$).

\subsection{Surface brightness}
\begin{figure}
\epsscale{0.8}
\plotone{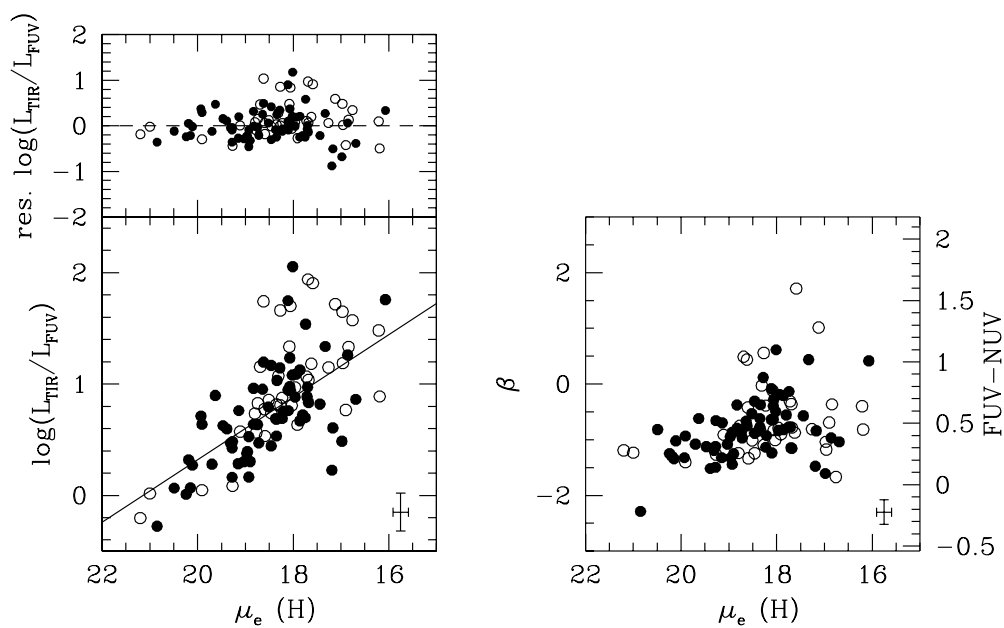}
\small{\caption{Relation between the mean H-band surface brightness ($\mu_{e}$) and the $L_{TIR}/L_{FUV}$ ratio (left) or $\beta$ (right). 
Symbols are as in Fig. \ref{IRX}.
The solid line shows the best linear fit for our \emph{primary} sample. The residuals from the best linear 
fit for normal galaxies are shown in the upper panel.}
\label{mue}}
\end{figure}

\begin{figure}
\epsscale{0.8}
\plotone{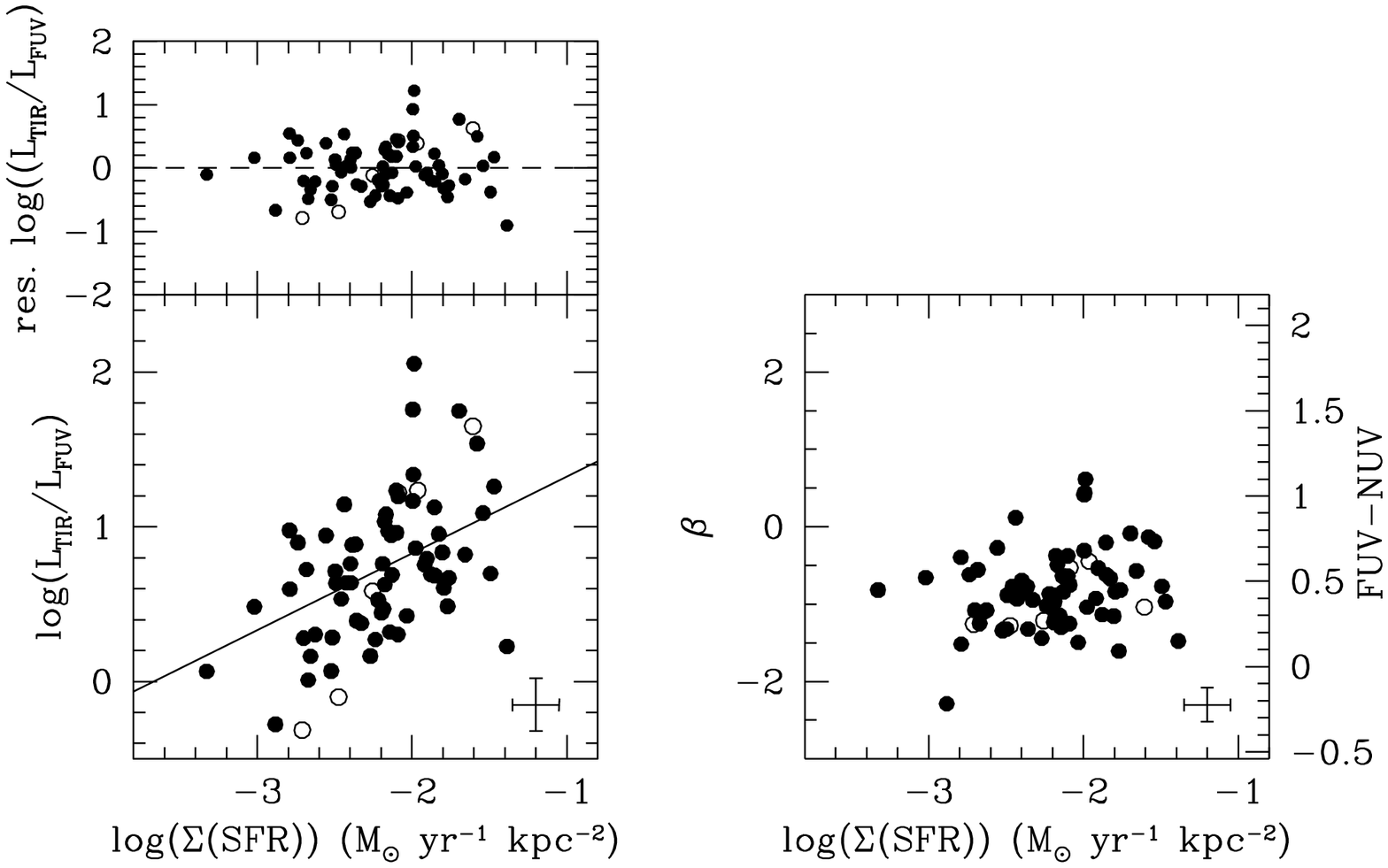}
\small{\caption{Relation between the star formation rate density and the $L_{TIR}/L_{FUV}$ ratio (left) or $\beta$ (right). 
Symbols are as in Fig. \ref{IRX}.
The solid line shows the best linear fit for our \emph{primary} sample.The residuals from the best linear 
fit for normal galaxies are shown in the upper panel.}
\label{muetot}}
\end{figure}

\cite{wang96} interpreted the increase of dust attenuation with rotational velocity (or mass) as due to the 
variations in both the metallicity and surface density of galactic disk with galactic size.
Fig.\ref{mue} shows the variation of the effective H-band surface 
brightness (defined as the mean surface brightness within the radius that contains 
half of the total galaxy light) 
and the dust attenuation. 
The two quantities are strongly anti-correlated ($r_{s}\sim$-0.63, $P(r_s)>$99.9\%): 
\begin{equation}
\log(\frac{L_{TIR}}{L_{FUV}}) = (-0.28\pm0.04) \times  \mu_{e}(H) + (5.92\pm0.81)
\end{equation}
with a scatter of $\sim0.34\pm0.03$ in $\log(L_{TIR}/L_{FUV})$: {\bf $\sim$1.2$\sigma$} lower than the value obtained 
for H-band luminosity and consistent with the one obtained for the gas metallicity.
Unfortunately in this case we cannot compare the behaviour of normal galaxies 
with the one of starbursts due to the lack of an estimate of $\mu_{e}$ for the starbursts.
Does this relation indicate that UV dust extinction depends on the thickness of stellar disk, 
or does it follows from the correlation between attenuation and star formation 
surface density?
To attack this question we determine the SFR density (defined 
as the ratio between the SFR determined from H$\alpha$ (eq.\ref{SFR}) and optical galaxy area).
Fig.\ref{muetot} shows the relation between the SFR 
density and $\log(L_{TIR}/L_{FUV})$.  
The two quantities are correlated ($r_{s}\sim$0.44, $P(r_s)>$99.9\%) with a dispersion 
of $\sim0.39\pm0.03$ in $\log(L_{TIR}/L_{FUV})$, 
$\sim$1.2$\sigma$ larger than the one observed for the mean H-band surface brightness\footnote{The difference between the two relations 
does not change if instead of the half-light radius, we use the total radius to estimate $\mu_{e}(H)$}.
Since the contribution of observational uncertainties to the scatter 
in the two relations is $\sim$ the same ($0.18\pm0.02$), our result might suggest that the UV attenuation is primarily correlated with 
the thickness of stellar disk, supporting the hypothesis of \cite{wang96} that both gas metallicity 
and star surface density are directly connected 
with the physical properties of dust (i.e. quantity and spatial distribution). 

\subsection{$L_{H\alpha}/L_{FUV}$ ratio}
\begin{figure}
\epsscale{0.8}
\plotone{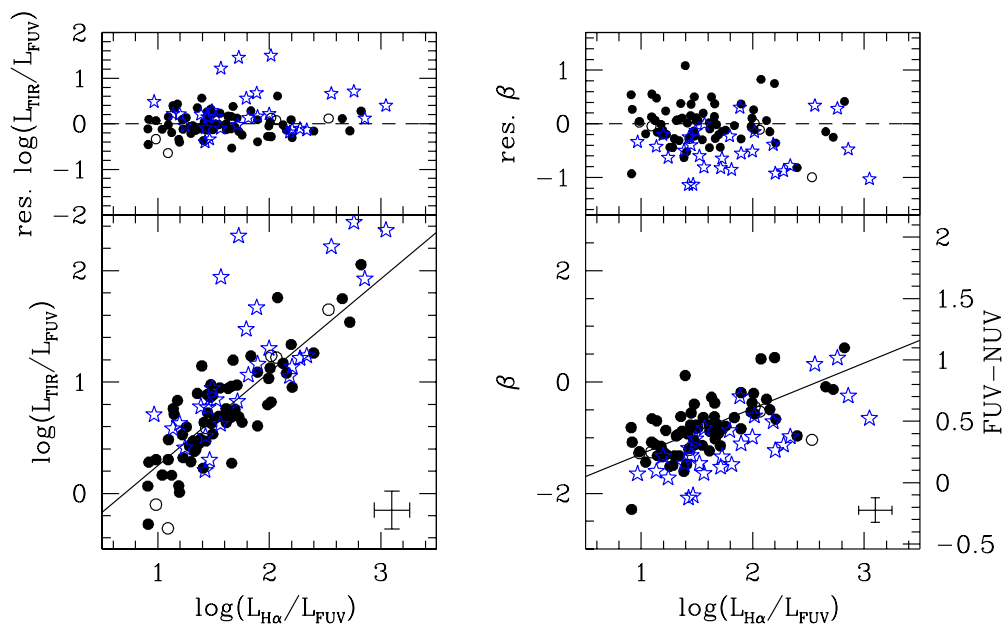}
\small{\caption{Relation between the H$\alpha$ and far ultraviolet luminosity and the $L_{TIR}/L_{FUV}$ ratio (left) or $\beta$ (right). Symbols are as in Fig.
\ref{IRX}. H$\alpha$ luminosity is corrected for dust attenuation using the Balmer decrement, while the FUV flux is uncorrected.
The solid lines show the best linear fit for our \emph{primary} sample. The residuals from the best linear 
fit for normal galaxies are shown in the upper panels.}
\label{hafuv}}
\end{figure}

\begin{figure}
\epsscale{0.8}
\plotone{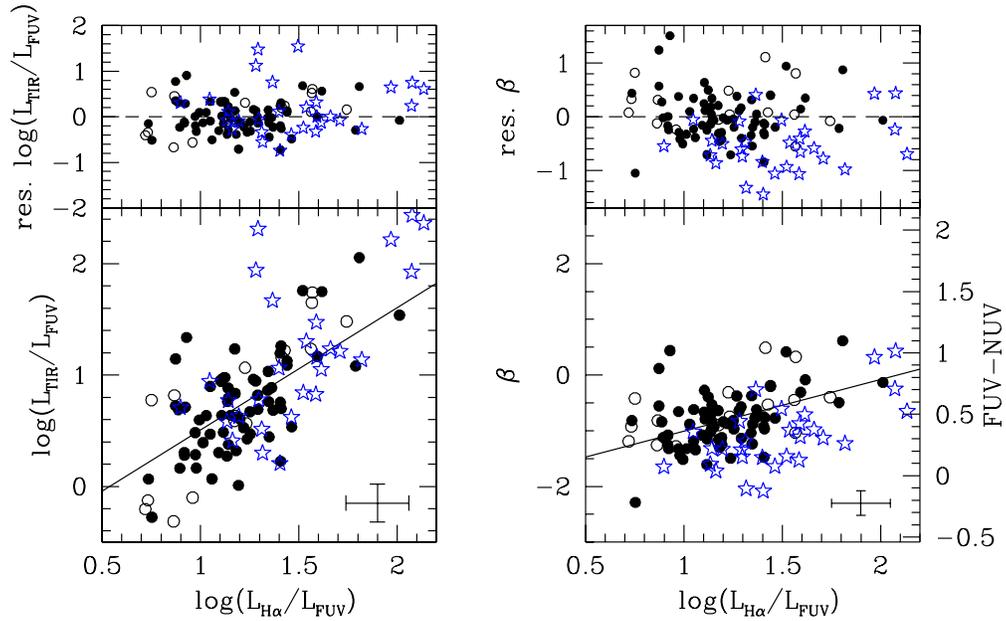}
\small{\caption{Relation between the observed H$\alpha$ and far ultraviolet luminosity and the $L_{TIR}/L_{FUV}$ ratio (left) or $\beta$ (right). Symbols are as in Fig.
\ref{IRX}. H$\alpha$ luminosity is the observed value not corrected for dust attenuation.
The solid lines show the best linear fit for our \emph{primary} sample.The residuals from the best linear 
fit for normal galaxies are shown in the upper panels.}
\label{hafuvnocor}}
\end{figure}

\cite{buat02} suggested that the $L_{H\alpha}/L_{FUV}$ ratio could be another potential attenuation indicator but 
they found a scattered correlation between  $L_{H\alpha}/L_{FUV}$ and $A(FUV)$, confirmed by \cite{bell02}.
This correlation is expected since both H$\alpha$ and UV emission are star formation indicators.
The H$\alpha$ luminosity comes from stars more massive than 10 M$_{\odot}$ and it traces the SFR 
in the last $\leq10^{7}$ yr while 
the UV luminosity comes from stars of lower mass (M$\geq1.5~ \rm M_{\odot}$) and it can be used 
as an indicator of the SFR in the last $\approx10^{8}$ yr.
This means that under the condition that the star formation is approximately constant 
in the last $\approx10^{8}$ yr the ratio $L_{H\alpha}/L_{FUV}$ 
(corrected for attenuation) should be fixed. Thus the ratio between the extinction 
corrected $L_{H\alpha}$ and the observed $L_{FUV}$ should be 
a potential attenuation indicator.
In Fig.\ref{hafuv} we analyze the relationship between the dust attenuation and the 
$L_{H\alpha}/L_{FUV}$ ratio, where $L_{H\alpha}$ 
is the H$\alpha$ luminosity corrected for dust attenuation using the Balmer decrement 
and for the contamination of $\rm[NII]$.
The two quantities turn out to be strongly correlated ($r_{s}\sim$0.76, $P(r_s)>$99.9\%):
\begin{equation}
\label{eqhafuv}
\log(\frac{L_{TIR}}{L_{FUV}}) = (0.84\pm0.07) \times \log(\frac{L_{H\alpha}}{L_{FUV}}) - (0.59\pm0.12)
\end{equation}
The dispersion around this relation is $\sim0.24\pm0.02$ in $\log(L_{TIR}/L_{FUV})$, 
consistent with the one 
observed for the $\log(L_{TIR}/L_{FUV})-\beta$ relation. 
The high correlation and low scatter between the two quantities 
is expected since the two variables are mutually related: the FUV luminosity appears 
in both axes and $L_{TIR}$ and $L_{H\alpha}$ are known to be correlated \citep{kewley02b}, 
explaining  why in the left panel of Fig.\ref{hafuv} starbursts and normal galaxies show the same trend.
The right-panel of Fig.\ref{hafuv} shows the relation between the ultraviolet slope 
and the $L_{H\alpha}/L_{FUV}$ ratio. 
In this case starbursts and normal galaxies behave differently: 
at any given $\beta$ starbursts have an higher 
$L_{H\alpha}/L_{FUV}$ than normal galaxies, consistent with 
what expected for galaxies experiencing a burst of star formation \citep{jorgehauv}.\\
A secure determination of the Balmer decrement for large samples is still a hard task, especially 
at high redshift, thus 
we look for a relation similar to Eq.(\ref{eqhafuv}) using the observed H$\alpha$ luminosity ($L_{H\alpha}^{obs}$).
The $L_{H\alpha}^{obs}/L_{FUV}$ and $\log(L_{TIR}/L_{FUV})$ ratios are 
yet correlated (see Fig.\ref{hafuvnocor}) but the correlation coefficient is lower than the previous case ($r_{s}\sim$0.49, $P(r_s)>$99.9\%).
The best linear fit gives:
\begin{equation}
\label{eqhafuvnoco}
\log(\frac{L_{TIR}}{L_{FUV}}) = (1.10\pm0.17) \times \log(\frac{L_{H\alpha}^{obs}}{L_{FUV}}) - (0.59\pm0.21)
\end{equation}
with a mean absolute deviation of $\sim0.34\pm0.03$ ($\sim$3.3$\sigma$ higher than for Eq.\ref{eqhafuvnoco}).
%We stress the reader that the use of Eq.(\ref{eqhafuv}) or  Eq.(\ref{eqhafuvnoco}) to estimate the amount of dust attenuation 
%will immediately invalid the use of the $L_{H\alpha}/L_{FUV}$ as an indicator of recent bursts of star formation \citep{jorgehauv}.}
%Since the $L_{H\alpha}/L_{FUV}$ is also 
%This result could be explained assuming that the $L_{H\alpha}/L_{FUV}$

\section{A cookbook for determining $L_{TIR}/L_{FUV}$ ratio in optically-selected galaxies}
\label{cook}
In this paper we investigated the relations between dust attenuation, 
traced by the $L_{TIR}/L_{FUV}$ ratio, and other global properties of normal star forming galaxies. 
Furthermore we compared the dust attenuation in normal and starbursts galaxies using multiwavelength datasets.
The amount of dust attenuation is found to correlate with the UV colors, gas metallicity, 
mass and mean surface density but, generally speaking, differently for normal and starburst galaxies.
Determine whether this difference is real or is due to aperture effects requires 
the analysis of GALEX observations for a sample of starburst galaxies.
The dispersion in the $L_{TIR}/L_{FUV}-\beta$ relation 
correlates with the birthrate parameter $b$,
suggesting that the observed scatter is, at least partly, due to differences in the star formation history. 
These results stress that estimating the UV dust attenuation, and consequently 
the star formation rate of normal galaxies (at high redshift in particular) is highly uncertain 
($\geq$50\%) 
when rest-frame far infrared observations are not available.
Moreover the sample selection criteria could strongly affect its properties, as 
recently pointed out by \cite{buat05} and \cite{denis05a}. 
They studied the dust attenuation properties and star formation activity in a UV and in a FIR
selected sample, showing that the former shows correlations 
with global galaxy properties, such as mass and bolometric luminosity, 
that the FIR selected sample does not.
Their results stress that the dust attenuation 
properties are very heterogeneous and that $L_{TIR}/L_{FUV}$ cannot be derived in a robust 
manner when far infrared observations are not available.\\
However the present investigation has shown that 
among optically-selected samples of normal galaxies with no nuclear activity a number of
empirical relations exists,
allowing to derive the $L_{TIR}/L_{FUV}$ ratio (and its uncertainty).
Once the attenuation at UV is determined it can be transformed to any other $\lambda$,  
only knowing the shape of the attenuation law and dust geometry 
(i.e. \citealp{calzetti94,gav02,bosellised}). \\  
In Table \ref{tablecook} we list all the relations, their associated  r.m.s., mean absolute 
deviation from the best fit (m.a.d.)\footnote{
The mean absolute deviation is less sensitive to the contribution of outliers than the standard deviation.  
For a Gaussian distribution the mean absolute deviation (m.a.d.) is $\sim\sqrt{2/\pi}\times(r.m.s.)$, while it is lower (higher) 
for a heavier (lighter) tailed distribution. As shown in Table 1 the values obtained for r.m.s. and m.a.d. are consistent 
with the ones expected for a Gaussian distribution} and 
the Spearman correlation coefficient.\\
Before we proceed describing our recipes, we have to investigate whether the scatter 
in these relations is physical or is only driven by  
observational uncertainties. In the latter case, in fact, our cookbook 
would not be very useful, since it would be valid only for observations with the 
same uncertainties than our datasets. 
For H-band luminosity, H-band surface brightness, $L_{H\alpha}^{obs}/L_{FUV}$ ratio and metallicity 
the contribution of observational uncertainties to the observed 
scatter varies from $\sim$ 18\% (r.m.s.$\sim0.17\pm0.02$) for $L_H$ to $\sim$40\% (r.m.s.$\sim0.21\pm0.02$) 
for $12+\log(O/H)$ and $L_{H\alpha}^{obs}/L_{FUV}$: even accounting for 
the contribution of measurements errors, the relative difference in the scatter of these relations 
does not change. On the contrary this confirms that the relation involving 
$L_H$ is the one with the highest "physical" dispersion, while for the other three relations the scatter is similar.\\  
The situation is worse for the relations involving $\beta$ and the $L_{H\alpha}/L_{FUV}$ ratio: 
the contribution of observational errors 
is $\sim$70-76\% ($\sim0.21\pm0.02$).
Thus it is impossible to determine which of these two relations 
has the lowest scatter and represents 
the best way to estimate dust attenuation without far infrared observations.
We can conclude that observational errors could account for the difference scatter observed in the relations 
involving $\beta$ and the $L_{H\alpha}/L_{FUV}$ ratio, but not for the difference observed 
in all the other relations. 
Our results can thus be used to suggest different ways to correct for UV dust attenuation.\\ 
%The logical flow-diagram in Fig.\ref{logicalflow} guides in their application.
Ia) The $L_{TIR}/L_{FUV}-\beta$ relation still represents one of the best 
way to quantify dust attenuation. The uncertainty in the value of 
$\log(L_{TIR}/L_{FUV})$ is $\sim0.26\pm0.03$.
%We looked for two ways to reduce the scatter on this relation: using the birthrate parameter or 
%the $EW(H\alpha+[NII])$. Unfortunately no statistically significative improvement in the scatter 
%is obtained (m.a.d.$\sim0.22\pm0.02$ and rms$\sim0.28\pm0.03$)}\\
%i) If both H-band and H$\alpha$ luminosity are available 
%the birthrate parameter can be computed using eq.(\ref{xuobs}) and the mean 
%uncertainty on $\log(L_{TIR}/L_{FUV})$ reduces to $\sim0.28$.\\
%ii) If only the $EW(H\alpha+[NII])$ is known the RMS can be reduced 
%to $\sim0.28$ using equation (\ref{xyew}).\\
%Otherwise:\\
Ib) If the UV spectral slope $\beta$ is unknown but we know $L_{H\alpha}$ (corrected for attenuation)  
we can obtain the ultraviolet attenuation using equation (\ref{eqhafuv}), with a r.m.s. of $0.24\pm0.02$.
This relation is valid under the assumption that the 
star formation rate is approximately constant 
in the last $\approx10^{8}$ yr. \\
%Obviously, as discussed in previous section, the corrected $L_{H\alpha}/L_{FUV}$ ratio so obtained 
%cannot be used to study the 
%recent star formation history of galaxies \citep{jorgehauv}; {\bf thus we suggest to use Eq.(\ref{eqhafuv}) 
%only when $\beta$ is unknown even if its dispersion is lower than the one observed 
%in the $L_{TIR}/L_{FUV}-\beta$ relation. 
%If we want to use the 
%$L_{H\alpha}/L_{FUV}$ ratio as an indicator 
%of recent bursts of star formation the best choices are, if available, H-band surface brightness or metallicity
%(m.a.d$\sim0.26\pm0.02$).\\
IIa) If we know $L_{H\alpha}^{obs}$, but no estimate of A(H$\alpha$) is available, 
we can use Eq.(\ref{eqhafuvnoco}) (rms$\sim0.34\pm0.03$).\\
IIb) If neither $\beta$ nor H$\alpha$ luminosity are available we are left with the relations 
with H-band surface brightness\footnote{Since we need H$\alpha$ flux to estimate metallicity, 
Eq.(\ref{relmet}) cannot be used in this case.} (r.m.s.$\sim0.34\pm0.03$) and, in the worse case,\\
III) H-band luminosity 
(rms$\sim0.39\pm0.03$ ).\\ 
%If more than one of these quantities are available we suggest to combine these methods.
%{\bf The m.a.d. reduces to $\sim0.22\pm0.02$ and the rms to $\sim0.32\pm0.03$ when metallicity, surface brightness and luminosity are available.}\\
Summarizing, these relations allow us 
to estimate the value of the $L_{TIR}/L_{FUV}$ ratio with an average uncertainties of$\sim$0.32 dex. 
This value corresponds approximately to $\sigma(A(FUV))\sim$0.5 mag, 
assuming $\log(L_{TIR}/L_{FUV})=1$ (the mean value for our sample)
and using the model of \cite{buat05}.
This is the lowest uncertainty on the estimate of the $L_{TIR}/L_{FUV}$ ratio in absence of 
far infrared observations. 
%, i.e. only a factor 2 higher than the mean observational 
%error ($\sim$0.16) in the estimate of the $\log(L_{TIR}/L_{FUV})$ from IRAS and GALEX observations. 
However we caution the reader that this value holds only for an optically-selected sample and that samples selected 
according to different criteria, especially FIR-selected, could contain higher dispersions.

\acknowledgements
We thank an unknown referee for her/his useful comments which helped us to improve and strengthen the paper. 
We wish to thank Christian Bonfanti, Jorge Iglesias-Paramo, Paolo Franzetti, Akio K. Inoue and Gerry Sanvito for useful
discussions. GALEX (Galaxy Evolution Explorer) is a NASA Small Explorer, launched in April 2003.
We gratefully acknowledge NASA's support for construction, operation,
and science analysis for the GALEX mission,
developed in cooperation with the Centre National d'Etudes Spatiales
of France and the Korean Ministry of Science and Technology. This research has made extensive use of the GOLDMine Database 
and of the NASA/IPAC Extragalactic Database (NED) 
which is operated by the Jet Propulsion Laboratory, California Institute of Technology, 
under contract with the National Aeronautics and Space Administration.
The authors would like to take this opportunity to thank the members
of the GALEX SODA Team for their valiant efforts in the timely
reduction of the complex observational dataset covering the full
expanse of the Virgo cluster.

\appendix
\section{Estimate of $A(H\alpha$)}
The attenuation in the Balmer lines %H$\alpha $ and H$\beta$ 
can be deduced from the comparison of the observed ratio $L_{H\alpha}/L_{H\beta}$ with the
theoretical value of 2.86 obtained for the recombination case B, an electronic density 
$n_{e}\leq10^{4}\rm ~cm^{-3}$ and temperature 
$\sim 10^{4}\rm ~K$. The variation of this value with density its negligible and with 
temperature is $\leq$5\% (in the range between 5000 K and 
20000 K, \citealp{caplan86}). 
The underlying absorption was deblended from the H$\beta$ emission line using a multiple component
fitting procedure. To do this the emission line is measured and subtracted from the spectra.
The resulting absorption line is also measured with respect to a reference continuum.
These two measurements are used as first guess in a fitting algorithm
which fits jointly the emission and absorption lines to the reference continuum.
For objects whose $H\beta$ is detected in emission
but the deblending procedure is not applied (no absorption feature is evident)
a mean additive correction for underlying absorption equal to -1.8 in flux and -1.4 \AA in EW is used.
These values correspond to the fraction of the (broader) absorption feature that lies under the emission line.
We adopt a dust screen geometry and the
Milky Way extinction curve (e.g. \citealp{kennicut83,calzetti94}).
Whereas varying the extinction curves has negligible effects in the visible,
the dust screen assumption seems to under-estimate the extinction by $\sim$0.2
mag compared with the amount deduced from the measurements of the thermal
radio continuum (\citealp{caplan86,bell01}).
We do not apply any correction for H$\alpha$ underlying absorption \citep{long01}.
However, since all the objects have $EW({\rm H\alpha}+[NII]) > 3\rm\AA$, the
underestimate in the value of $A(H\alpha$) is negligible.
In fact no change (at a 99\% significance level) is 
observed comparing the best fits obtained in this work and 
the ones obtained adding to the H$\alpha$ the same fixed underlying absorption used 
for H$\beta$ when the underlying is not detected.
We assume that the errors on $A(H\alpha $) are mainly due to the uncertainty
on the H$\beta$ flux. These
errors represent in fact the lower limits because we do not account for the
uncertainty introduced by the fitting of the lines.
They range from 0.01 to 0.43 mag and are found strongly
anti-correlated with EW(H$\beta$) (see \citealp{gavspectra}).
Adopting the definition of the Balmer decrement as in \cite{gavspectra}:
\begin{equation}
\label{C1}
C1(H\beta) = \frac{log(\frac{1}{2.86} \times \frac{L_{H\alpha}}{L_{H\beta}})}{0.33}
\end{equation}
Since the $A(H\alpha)$ attenuation is:
\begin{equation}
\label{aha}
A(H\alpha)= 1.086 \frac{1}{e_{\beta\alpha}-1}ln(\frac{1}{2.86} \times \frac{L_{H\alpha}}{L_{H\beta}})
\end{equation}
From (\ref{C1}) and (\ref{aha}) we obtain:
\begin{equation}
A(H\alpha)= 1.086 \frac{1}{e_{\beta\alpha}-1} \times 0.33 \times C1(H\beta) ~ln(10)
\end{equation}
and assuming a galactic extinction law ($e_{\beta\alpha}$ = 1.47) we derive:
\begin{equation}
A(H\alpha)= 1.756 \times C1(H\beta)
\end{equation}
$A(H\alpha $) = 0.85 mag is obtained on average, consistent with
previous studies (e.g. \citealp{kennicut83,kennicut92,thuan92,kewley02b}).
Eleven galaxies have H$\beta$ undetected in emission but the underlying
stellar absorption is clearly detected. For them we derive a $3 \times \sigma$lower 
limit to the H$\beta$ flux ($f_{H\beta}$) using \citep{gavspectra}:
\begin{equation}
f_{H\beta}<3\times  rms_{(4500-4800)} \times H\alpha(HWHM)
\end{equation}
assuming that $H\alpha $ and $H\beta $ emission lines have similar HWHM (Half Width Half Maximum).
As shown in Eq.(\ref{C1}) a change in the theoretical value of the $L_{H\alpha}/L_{H\beta}$ ratio 
would only produce a small ($\leq$5\%) constant over (or under) estimate of the ionized gas attenuation, thus leaving 
unchanged the shape and dispersions of the observed relations, only affecting 
the values of the best fitting parameters.

\section{Dust to Gas ratio}
\begin{figure}
\epsscale{0.8}
\plotone{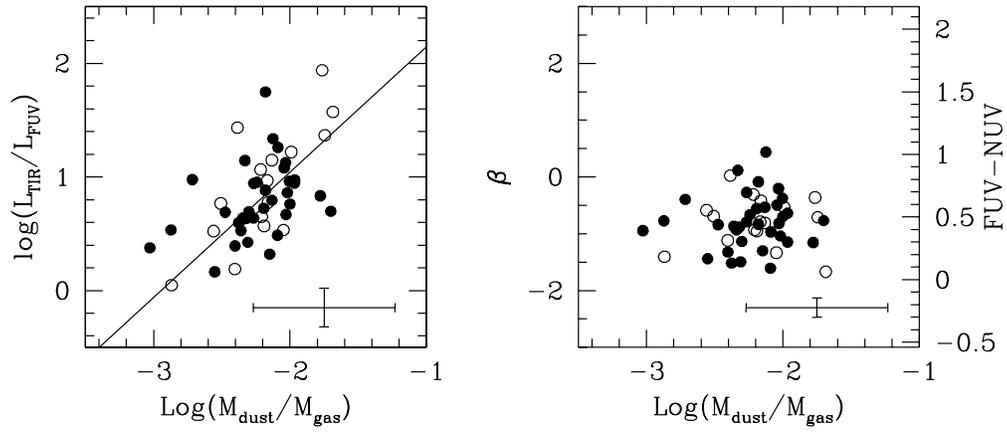}
\small{\caption{Relation between the gas to dust ratio and the $L_{TIR}/L_{FUV}$ ratio (left) or $\beta$ (right). Symbols are as in Fig. \ref{IRX}.
The solid line shows the best linear fit for our \emph{primary} sample.}
\label{gdust}}
\end{figure}

The correlation between attenuation and metallicity can be interpreted assuming
that the ultraviolet radiation produced by star forming regions suffers a dust 
attenuation increasing with the dust to gas ratio, which correlates with metallicity.
(e.g. \citealp{issa90,inoue03}).
In order to  check this hypothesis we compute the dust to gas ratio following \cite{boselligdust}.
In normal galaxies the dust mass is dominated by the cold dust emitting above $\sim$200 $\mu $m.
The total dust mass can be estimated
provided that the 100-1000 $\mu $m far-IR flux and the cold dust temperature are known.
Fitting the SEDs of normal galaxies with a modified
Planck law $\nu$$^{\beta}$ $B_{\nu}(T_{\rm d})$, with $\beta=2$ \citep{alton00},
the total dust mass can be determined from the relation \citep{devereux90}:
\begin{equation}
\label{dustmass}
{M_{\rm dust}=CS_{\lambda}D^2({\rm e}^{a/T_{\rm dust}}-1)~{M_\odot}}
\end{equation}
where $C$ depends on the grain opacity, $S_{\lambda}$is the far-IR flux at a given wavelength (in Jy),
 $D$ is the distance of the galaxy (in Mpc), $T_{\rm dust}$
 is the dust temperature, and $a$  depends on $\lambda$.
 Only IRAS data at 60 and 100 $\mu $m are available for our sample and,
 given the strong contamination of the emission at 60 $\mu $m by very
 small grains, the 60 to 100 $\mu $m ratio does not provide a reliable measure of $T_{\rm dust}$
 \citep{contursi01}. $T_{\rm dust}$ determined by
\cite{alton98} consistently with \cite{contursi01}, seems  to be independent of the UV radiation field, 
 of the metallicity or of the total luminosity \citep{boselligdust}.
  Therefore we will adopt the average value $T_{\rm dust}=20.8\,\pm\, 3.2$ K for all our galaxies
 introducing an uncertainty of $\sim$50\% on the estimate of $M_{dust}$ 
 (equation (\ref{dustmass})). 
 We then estimate
 the dust mass of the sample galaxies using (\ref{dustmass})  with
 $C=1.27~\rm M_{\odot} Jy^{-1} Mpc^{-2}$, consistent with \cite{contursi01},
 and $a$=144 K for $S_{\lambda}=S_{\rm 100\,\mu m}$ \citep{devereux90}.
The determination of the dust to gas ratio, in a way consistent with
that obtained in the solar neighbourhood, requires the estimate of
the gas and dust surface densities, thus of the spatial distribution
of dust and gas over the discs.
Unfortunately only integrated HI and $\rm H_{2}$ masses are available
for our spatially unresolved galaxies.
It is however reasonable to assume that the cold dust and
the molecular hydrogen are as extended as the optical disc \citep{alton98,boselligdust}.
To determine the mean HI surface density we adopt (\cite{boselligdust}):
\begin{displaymath}
\log \Sigma_{\rm HI} = 20.92 (\pm 0.17) - 0.65 (\pm 0.11) \times (def(HI))~\rm {cm^{-2}}
\end{displaymath}
where def(HI) is the galaxy HI deficiency.
Thus the dust to gas ratio is obtained from the ratio of the dust surface density to the sum of 
molecular and neutral hydrogen surface densities. 
In Fig. \ref{gdust} we compare the relation between the $L_{TIR}/L_{FUV}$ ratio (left panel) and $\beta$ (right panel)
with the dust to gas ratio. The gas to dust ratio barely correlates with the $L_{TIR}/L_{FUV}$ ratio (R$\sim$0.38).
Contrary to metallicity, we do not find a significant 
correlation (R$\sim$0.11) with the ultraviolet spectral slope.
This is probably due to the high uncertainty in our estimate of $M_{dust}$ consequent 
to assuming the same temperature for all our galaxies ($M_{dust} \propto e^{a/T_{dust}}$, thus 
small errors ($\sim$15\%) on $T_{dust}$ propagate onto $\sim$50\% errors on $M_{dust}$).

\clearpage

\begin{table}
\caption {Linear realtions useful to estimate the $L_{TIR}/L_{FUV}$ ratio ($\log(L_{TIR}/L_{FUV})=a\times x+b$).}
\[
\label{tablecook}
\begin{array}{p{0.17\linewidth}ccccc}
\hline
\noalign{\smallskip}
$x$ & a & b & m.a.d.\tablenotemark{a} & rms\tablenotemark{b} & r_{s} \\
\noalign{\smallskip}
\hline
\noalign{\smallskip}									       
$\beta$ 	             & 0.70\pm0.06   &   1.30\pm0.06  &  0.20\pm0.02 & 0.26\pm0.02 & 0.76 \\
%$\beta,b$   & (1.04\pm0.09)\times\beta + (0.35\pm0.11)\times\log(b) + (1.77 \pm0.11)   &0.28	  \\
%$\beta,EW(H\alpha)$  & (1.04\pm0.09)\times\beta + (0.34\pm0.11)\times EW(H\alpha+[NII]) + (1.05 \pm0.19)  &0.28	  \\
%$12+\log(O/H)$               & 2.15\pm0.27   & -18.17\pm1.50   &  0.30\pm0.03 & 0.38\pm0.03 & 0.59  \\
$12+\log(O/H)$               & 1.37\pm0.24   & -11.36\pm2.11   &  0.26\pm0.02 & 0.35\pm0.03 & 0.59  \\
%$L_{H}/L_{\odot}$	     & 0.83\pm0.16   & -7.55\pm0.58    &  0.35\pm0.04 & 0.46\pm0.04 & 0.49  \\
$L_{H}/L_{\odot}$	     & 0.34\pm0.10   & -2.66\pm0.88    &  0.29\pm0.03 & 0.39\pm0.03 & 0.49  \\
$\mu_{e}(H)$	             & -0.28\pm0.04  & 5.92\pm0.81    &  0.25\pm0.02 & 0.34\pm0.03  & -0.63 \\
%$\mu_{e}(H)$	             & -0.48\pm0.06  & 9.61\pm0.44    &  0.29\pm0.03 & 0.38\pm0.03  & -0.63 \\
%$L_{H\alpha}/L_{FUV}$	     & 1.01\pm0.06   &-0.86\pm0.07    &  0.23\pm0.02 & 0.30\pm0.03  & 0.76  \\
$L_{H\alpha}/L_{FUV}$	     & 0.84\pm0.07   &-0.59\pm0.12    &  0.19\pm0.02 & 0.24\pm0.02  & 0.76  \\
$L_{H\alpha}^{obs}/L_{FUV}$  & 1.10\pm0.17   &-0.59\pm0.21    &  0.27\pm0.02 & 0.34\pm0.03 & 0.49  \\

\noalign{\smallskip}
\hline
\end{array}
%\tablenotetext{a}{All the fit are performed using a unweighted bisector linear fit procedure.}
\tablenotetext{a}{Mean absolute deviation from the best fit}
\tablenotetext{b}{Standard deviation from the best fit}
\]
\end{table}

\clearpage

\clearpage

\clearpage

%\begin{figure}
%\epsscale{1.0}
%\plotone{f3.ps}
%\small{\caption{Relation between the uncorrected (left) and corrected (right) FUV-H color and the distance $\Delta <(L_{TIR}/L_{FUV})-\beta>$ from
%the $L_{TIR}/L_{FUV}-\beta$ relation for our sample. The {\bf solid line} in the right panel indicates 
%the best linear fit to our data.}
%\label{resFUV}}
%\end{figure}

\clearpage

\clearpage

\clearpage

\clearpage

\clearpage

\clearpage

\clearpage

\clearpage

\clearpage

\clearpage

\end{document}